\documentclass[twocolumn,aps,prl] {revtex4} 
\usepackage{graphicx}
\begin{document}
\pagestyle{empty} 
\title{Contact mechanics: contact area and interfacial separation from small contact to full contact}
\author{C. Yang\footnote{ Email address: c.yang@fz-juelich.de} and B.N.J. Persson}
\affiliation{Institut f\"ur Festk\"orperforschung, Forschungszentrum J\"ulich, D-52425 J\"ulich, Germany}

\begin{abstract}
We present a molecular dynamics study of
the contact between a rigid solid with a randomly rough surface
and an elastic block with a flat surface. The numerical calculations  mainly focus on the contact area
and the interfacial separation from small contact (low load) to full
contact (high load). 
For small load the contact area varies linearly with the load and the interfacial separation
depends logarithmically on the load. For high load the contact area approaches the nominal contact area
(i.e., complete contact), and the interfacial separation approaches zero.
The numerical results have been compared with analytical theory and experimental results.
They are in good agreement with each other. The present findings may be
very important for soft solids, e.g., rubber, or for very smooth surfaces, where complete contact
can be reached at moderate high loads without plastic deformation of the solids. 
\end{abstract}
\maketitle


{\bf 1. Introduction}

What happens at the atomic and molecular level when surfaces come into contact with each other?
And how do these events relate to macroscopic properties and observations? These questions, which
center on the phenomena of adhesion and friction, pose challenges not only in engineering but also
in many areas of physical and biological sciences\cite{skimming}. Finding correlations 
and models that connect the atomic and macroscopic worlds usually is not easy. However, 
recently we have surprisingly found that the pressure distribution obtained from 
molecular dynamics calculations
is in a good agreement with the prediction based on continuum contact mechanics, and in 
particular with the analytical contact mechanics theory of Persson\cite{JCPpers,P1,preparation}.

When two elastic solids with rough surfaces are squeezed together, the solids will
in general not make contact everywhere in the apparent contact area,
but only at a distribution of asperity contact spots\cite{Borri,Hyun1}. The separation
$u({\bf x})$ between the surfaces will vary in a nearly random way with the lateral
coordinate ${\bf x}=(x,y)$ in the apparent contact area. 
When the applied squeezing pressure increases, the contact area $A$ will increase and the average 
surface separation $\bar{u}=\langle u({\bf x})\rangle$ will decrease, 
but in most situations it is not
possible to squeeze the solids into perfect contact corresponding to $\bar u=0$. 
The area of real contact, and the space between
two solids has a tremendous influence on many important processes.

Most studies of contact mechanics have been focused on small load where the contact area
depends linearly on the load\cite{Bush,GW, Hyun3, Chunyan,Carlos, Popov}. 
However, for soft solids, such as rubber or gelatin, and for smooth surfaces
nearly full contact may occur at the interface, and it is of great interest to study how the
contact area, the interfacial surface separation and  stress distribution vary with
load from small load (where the contact area varies linearly with the load),
to high load (where the contact is (nearly) complete). Here we will present such a study using
molecular dynamics, and we will compare the numerical results with the prediction of the analytical
contact mechanics theory of Persson.
Our multiscale molecular dynamics approach {\cite{Chunyan}} has been developed to study contact mechanics 
for surfaces with roughness on many different length scales,
e.g., self-affine fractal surfaces.

This paper presents an extension of the work 
reported in two short publications
\cite{preparation,MD.PRL}. 
In Sec. 2 we briefly review the contact mechanics of Persson.
In Sec. 3 we consider the 
relation between interfacial separation and squeezing pressure. Sec. 4 deals with the
molecular dynamics (MD) model. In Sec. 5 and 6 we compare the numerical results of the MD model with the
analytical theory for the real contact area
and interfacial separation, respectively. In Sec. 7 we compare
Persson theory  with finite element calculations. Sec. 8 deals with how average surface
separation depends on the magnification $\zeta$. In Sec. 9 we consider the adhesion between
randomly rough surfaces. Sec. 10 contains the summary and conclusion.

\begin{figure}
\includegraphics[width=0.45\textwidth,angle=0]{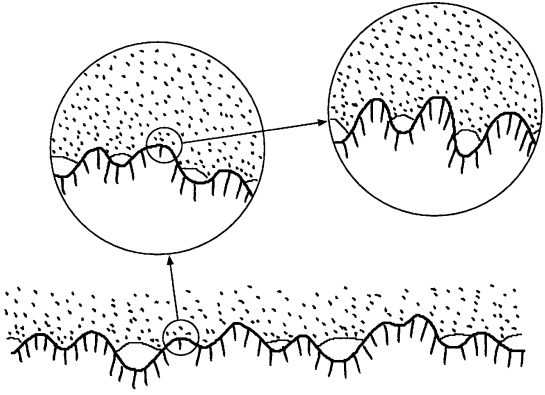}
\caption{\label{1x}
An rubber block (dotted area) in adhesive contact with a hard
rough substrate (dashed area). The substrate has roughness on many different 
length scales and the rubber makes partial contact with the substrate on all length scales. 
When a contact area 
is studied at low magnification it appears as if complete contact occur, 
but when the magnification is increased it is observed that in reality only partial
contact occur.  
}
\end{figure}

\vskip 0.3cm

{\bf 2. Theory: Contact area}

We consider the frictionless contact between elastic solids with randomly rough surfaces.
If $z=h_1({\bf x})$ and $h_2({\bf x})$ describe the surface profiles, $E_1$ and $E_2$ are
the Young's elastic moduli of the two solids and $\nu_1$ and $\nu_2$ the corresponding
Poisson ratios, then the elastic contact problem is equivalent to the contact between a
rigid solid (substrate) with the roughness profile $h({\bf x}) = h_1({\bf x})+h_2({\bf x})$, in contact
with an elastic solid (block) with a flat surface and with an Young's modulus $E$ and Poisson ratio
$\nu$ chosen so that\cite{Johnson1,Johnson2}
$${1-\nu^2 \over E} = 
{1-\nu_1^2 \over E_1} + 
{1-\nu_2^2 \over E_2}.$$ 

Persson\cite{JCPpers,P1} has developed a contact mechanics
theory where the surfaces are studied at different
magnification $\zeta = \lambda_0/\lambda$, where $\lambda_0$ is some reference length, e.g., the 
roll-off wavelength of the surface roughness power spectra (see below), and $\lambda$
the shortest wavelength roughness which can be observed at the magnification 
$\zeta$, see Fig. \ref{1x}. We define $q_0 = 2 \pi /\lambda_0$.
In this theory\cite{JCPpers} the stress distribution
$P(\sigma, \zeta)$ at the interface between the
block and the substrate has been shown to obey (approximately) a diffusion-like
equation where time is replaced by magnification and spatial coordinate by the 
stress $\sigma$.
When the magnification is so small that no atomic structure can be detected, 
the surface roughness will be smooth (no abrupt or step-like changes in the 
height profile) and one can
then show\cite{Persson_PRB_2002} that in the absence of
adhesion $P(0,\zeta)=0$. Using this boundary condition
the solution to the diffusion-like equation gives the pressure distribution
at the 
interface ($\sigma > 0$):
\begin{equation}
 \label{pressuredist}
 P(\sigma, \zeta) = {1\over 2 (\pi G)^{1/2}} 
\left (e^{-(\sigma -p)^2/4G}-e^{-(\sigma +p)^2/4G}\right ),
\end{equation}
where $p$ is the nominal squeezing pressure, and where
\begin{equation}
 \label{G}
 G= {\pi \over 4} \left ( {E\over 1 - \nu^2}\right )^2 \int_{q_L}^{\zeta q_0}dq \ q^3 C(q) \,.
\end{equation}
where $q_L$ is the smallest surface roughness wave-vector which may be of order
$ 2 \pi /L$, where $L$ is the linear size of the system.
The surface roughness power spectrum\cite{P3}
$$C(q)= {1\over (2\pi )^2} \int d^2x \langle h({\bf x}h({\bf 0})\rangle
e^{-i{\bf q}\cdot {\bf x}},$$
where $\langle..\rangle$ stands for ensemble average.  
The relative contact area
\begin{equation}
 \label{relatcontact1}
 {A\over A_0} = \int_{0^+}^\infty  d\sigma \ P(\sigma,\zeta)\equiv P(q),
\end{equation}
where $q=\zeta q_{0}$.
Sometimes, when we want to emphasize
that $P(q)$ depends on the pressure $p$, we will denote it by $P_{p}(q)$.
Note that there is a delta-function contribution to $P(\sigma, \zeta)$ 
of the form $[(A_0-A)/A_0]\delta (\sigma)$ which arises from the non-contact area. Including this 
delta-function the integral of $P(\sigma, \zeta)$ over all $\sigma$ will be unity as expected for a 
probability distribution.
In what follows we will always consider $\sigma > 0$ in which case $P(\sigma, \zeta)$ is given
by (1).  
Substituting (\ref{pressuredist}) into (\ref{relatcontact1}) gives after some simplifications
\begin{equation}
 \label{relatcontact2}
 {A\over A_0} =
 {1\over (\pi G)^{1/2}} \int_0^{p} d\sigma \ e^{-\sigma^2/4G}
  = {\rm erf}\left ({p \over{2 G^{1/2}}}\right ) .
\end{equation}
Thus, for small nominal squeezing pressure $p \ll G^{1/2}$ we get
\begin{equation}
 \label{relatcontact3}
 {A\over A_0} \approx {p \over (\pi G)^{1/2}} \,.
\end{equation}

\begin{figure}
\includegraphics[width=0.45\textwidth,angle=0]{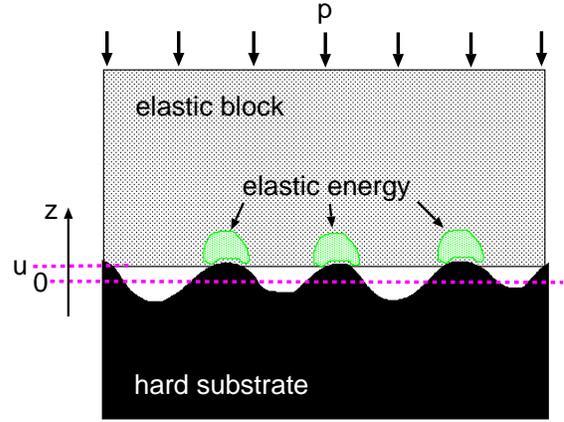}
\caption{\label{upimage}
An elastic block squeezed against a rigid rough substrate. The separation
between the average plane of the substrate and the average plane of the lower
surface of the block is denoted by $u$. Elastic energy is stored in the block in the vicinity
of the asperity contact regions.
}
\end{figure}

A critical discussion of the theory presented above was given by Manners and Greenwood
\cite{Greenwood}. See also Carbone and Bottiglione\cite{Giuseppe}.

\vskip 0.3cm

{\bf 3. Theory: Interfacial surface separation}

The space between
two solids has a tremendous influence on many important processes, e.g., heat transfer\cite{heat},
contact resistivity\cite{Rab}, lubrication\cite{BookP}, sealing\cite{sealing} and
optical interference\cite{Benz}. 
One of us has recently presented a simple
theory for the (average) separation $\bar{u}$ as a function of the squeezing pressure $p$ \cite{preparation}.
The theory shows that for randomly rough surfaces at low squeezing pressures $p \propto {\rm exp} (-\bar{u}/u_0)$
where the reference length $u_0$ depends on the nature of the surface roughness but is independent of $p$,
in good agreement with experiments\cite{Benz}.

Consider an elastic block with a flat surface squeezed against a hard rough substrate surface,
see fig. \ref{upimage}.
The separation between the average surface plane of the block and the
average surface plane of the substrate is denoted by $\bar{u}$ with $\bar{u} \ge 0$.
When the applied squeezing force $p$
increases, the separation between the surfaces at the interface will decrease, and
we can consider $p=p(\bar{u})$ as a function of $\bar{u}$. 
The elastic energy $U_{\rm el}(\bar{u})$ stored in the substrate asperity--elastic block contact regions
must equal the work done by the external pressure $p$ in displacing the lower surface of the
block towards the
substrate, i.e.,
\begin{equation}
   \label{equal}
    \int_{\bar{u}}^\infty du' A_0 p(u') = U_{\rm el}(\bar{u}), 
\end{equation}
or
\begin{equation}
   \label{simple_eq}
   p(\bar{u}) = - {1\over A_0} {d U_{\rm el} \over d\bar{u}},
\end{equation}
where $A_0$ is the nominal contact area. Equation (\ref{simple_eq}) is exact, and shows that if the
dependence of the surface separation $u$ on the squeezing pressure $p$ is known, 
e.g., from finite element calculations or
molecular dynamics, one can obtain the elastic energy
$U_{\rm el}$ stored in the asperity contact regions\cite{Chunyan}. 
This is an important result as $U_{\rm el}(\bar{u})$ is relevant for 
many important applications.

Theory shows that for low squeezing pressure, the area of real contact $A$ varies linearly with
the squeezing force $pA_0$, and that the interfacial stress distribution, and the 
size-distribution of contact spots, are independent of the squeezing pressure\cite{Arch,PSSR}. 
That is, with increasing $p$
existing contact areas grow and new contact areas form in such a way that in the thermodynamic limit
(infinite-sized system) the quantities referred to above remain unchanged. It follows immediately
that for small load {\it the elastic energy stored in the asperity contact region will increase
linearly with the load}, i.e., $U_{\rm el}(\bar{u}) = u_0 A_0 p(\bar{u})$, where $u_0$ is a characteristic length
which depends on the surface roughness
(see below) but is independent of the squeezing pressure $p$. Thus, for small pressures (7) takes the form
$$p(\bar{u}) = - u_0 {d p \over d\bar{u}},$$
or
\begin{equation}
  \label{exponential}
  p(\bar{u}) \propto e^{- \bar{u}/u_0},
\end{equation}
in good agreement with experimental data for the contact between elastic solids when the
adhesional interaction between the solids can be neglected\cite{Benz}. 
We note that the result (8) differs drastically from
the prediction of the Bush et al theory\cite{Bush}, and that of Greenwood and Williamson theory (GW)\cite{GW}, 
which for low squeezing pressures (for randomly rough
surfaces with Gaussian height distribution) predict $p(\bar{u}) \propto \bar{u}^{-a} {\rm exp} (-b\bar{u}^2)$, where $a=1$ in the
Bush et al theory and $a=5/2$ in the GW theory. Thus {\it these theories do not correctly describe the
interfacial spacing between contacting solids}. 
This is not surprising because these approaches assume a rigid substrate surface 
covered with flexible asperities.
In reality, the bulk of the solids whose surfaces are in contact is not rigid. Furthermore, there
exist a hierarchy of asperities on many length scales, all of which can distort.

The elastic energy $U_{\rm el}$ has been studied in Ref. \cite{P1} and \cite{PSSR}.
Here we will use 

\begin{equation}
  \label{old4}
  U_{\rm el} \approx A_0 E^* {\pi \over 2} \int_{q_0}^{q_1} dq \ q^2W(q,p)C(q),
\end{equation}
where we have chosen $q_0=q_L$ and
where $E^*=E/(1-\nu^2)$, $q_1$ is the largest surface roughness wave vector and
\begin{equation}
  \label{old77}
  W(q,p) = P_p(q) \left [\gamma +(1-\gamma) P_p^2(q)\right ],
\end{equation}
where $P_p(q)$ is 
given by Eq. (\ref{relatcontact2}):
\begin{equation}
  \label{errorfunction}
  P_p(q)= {\rm erf}\left ({p\over 2 G^{1/2}(\zeta)}\right ).
\end{equation}
For complete contact (infinite squeezing pressure) $P_p= 1$ and thus $W(q,p)=1$ and in this limit
(\ref{old4}) is exact. For small squeezing pressure $W(q,p) \approx \gamma P_p$.
The parameter $\gamma$ is of order $ \approx 0.4$ (see below),
and takes into account that the elastic energy stored in the contact
region (per unit surface area) in general is less than the 
average elastic energy (per unit surface area) for perfect contact (see Ref. \cite{PSSR}).
The particular way we interpolate between the limits $W(q,p) = 1$ for complete contact and
$W(q,p) = \gamma P_p$ for very small contact using (\ref{old77}) was designed to give good
agreement between the calculated [using (7)] interfacial separation, and the interfacial
separation obtained from Molecular Dynamics (MD) 
and Finite Element Method (FEM) (see Sec. 6 and 7).

Let us write (11) as\cite{JCPpers,Bucher} 

\begin{equation}
   \label{old5}
   P_p(q) = {2\over \surd \pi} \int_0^{s(q)p} dx \ e^{-x^2},
\end{equation}
where $s(q)=w(q)/E^*$ with 
$$w(q)=\left (\pi \int_{q_0}^q dq' \ q'^3 C(q') \right )^{-1/2}.$$
Using (12) gives
\begin{equation}
  \label{old7}
  {\partial P_p \over \partial \bar u} = {2\over \surd \pi} s {d p \over d \bar u} e^{-s^2p^2},
\end{equation}
Substituting (\ref{old4}) and (\ref{old77}) in (\ref{simple_eq}), and using (\ref{old7}) gives
$$p(\bar{u})=-\surd \pi \int_{q_0}^{q_1} dq \ q^2C(q) w(q) \left [\gamma+3(1-\gamma)P_{p}^{2}(q)\right ]$$

$$\times e^{-[w(q) p/E^*]^2} {d p \over d \bar{u}}, $$ 
or
$$d\bar{u} =-\surd \pi \int_{q_0}^{q_1} dq \ q^2C(q) w(q)\left [\gamma+3(1-\gamma)P_{p}^{2}(q)\right ]$$  
$$\times e^{-[w(q) p/E^*]^2} {d p \over p}. $$
Integrating this from $\bar u=0$ (complete contact, corresponding to $p=\infty$) to $\bar u$ gives
$$\bar u= \surd \pi \int_{q_0}^{q_1} dq \ q^2C(q)w(q)$$
\begin{equation}
   \label{old8}
\times \int_p^\infty dp' \ {1 \over p'} 
\left [\gamma+3(1-\gamma)P_{p'}^2(q)\right ] e^{-[w(q) p'/E^*]^2}.
\end{equation}

Let us consider the limiting case of very low squeezing pressure. If we introduce
$x=w(q)p'/E^*$ the last integral in (\ref{old8}) becomes
$$J=\int_{pw(q)/E^*}^\infty dx \ {1 \over x} 
\left [\gamma+3(1-\gamma)P^2(x)\right ] e^{-x^2},$$
where
$$P(x)= {2\over \surd \pi} \int_0^x dx' \ e^{-x'^2}.$$
Performing a partial integration gives
$$J= \left [ {\rm log} x \ \left [\gamma+3(1-\gamma)P^2(x)\right ] e^{-x^2} \right ]_{pw(q)/E^*}^\infty$$
$$-\int_{pw(q)/E^*}^\infty dx \ {\rm log} x  \ \Big ( 6(1-\gamma)P(x) P'(x)$$ 
$$+ \left [\gamma+3(1-\gamma)P^2(x)\right ] (-2x)\Big ) e^{-x^2}.$$
The leading contributions to $J$ as $p\rightarrow 0$ is
$$J= -\gamma {\rm log} \left ({pw(q) \over E^*}\right ) 
-\int_0^\infty dx \ {\rm log} x  \ \Big ( 6(1-\gamma)P(x) P'(x)$$ 
$$+ \left [\gamma+3(1-\gamma)P^2(x)\right ] (-2x)\Big ) e^{-x^2} 
= -\gamma {\rm log} \left ({pw(q) \over \epsilon E^*}\right ).\eqno(15)$$
where
$$\epsilon={\rm exp}\Bigg [ -\int_0^\infty dx \ {\rm log} x  \ \Big ( 6{1-\gamma\over \gamma}P(x) P'(x)$$ 
$$+ \left [1+3{1-\gamma\over \gamma} P^2(x)\right ] (-2x)\Big ) e^{-x^2}\Bigg ].\eqno(16)$$
Using (16) we get $\epsilon = 4.047$. Note that $\epsilon$ depends on $P_p(q)$ for all pressures
from small relative contact area ($P << 1$) to complete contact corresponding to $P=1$.
Thus, although the slope of the linear relation between $u$ and ${\rm log} p$, which hold for very
small $p$, only depend on $P_p(q)$ for very small $p$ (where the relative contact area is proportional
to $p$), the lateral position of the line {\it does} depend on the whole function
$P_p(q)$ [or $P(x)$]. For this reason it is important to accurately describe how
$U_{\rm el}$ depend on $P_p(q)$ for all $p$, even if one is only interested in the relation between
$\bar u$ and $p$ for very small $p$. The physical reason for this is simple: 
even for arbitrary small applied nominal stress $p$,
the stress (at high enough magnification) in the area of contact will be very high, 
which may result in (nearly)
complete contact in the asperity contact regions.

Substituting (15) in (14) gives
$$\bar{u}= -\surd \pi \int_{q_0}^{q_1} dq \ q^2C(q)w(q)
\gamma {\rm log} \left ({pw(q) \over \epsilon E^*}\right ) $$
$$= -u_0 [{\rm log} p - {\rm log} (\beta E^*)],$$
or
$$p=\beta E^* e^{-\bar{u}/u_0},\eqno(17)$$
where 
$$u_0= \surd \pi \gamma \int_{q_0}^{q_1} dq \ q^2C(q)w(q),\eqno(18)$$
and
$$\beta = \epsilon {\rm exp} \left [-{\int_{q_0}^{q_1} dq \ q^2C(q)w(q) {\rm log} [w(q)] \over 
\int_{q_0}^{q_1} dq \ q^2C(q)w(q)}\right ].\eqno(19)$$

The relation (\ref{exponential}) between $p$ and $\bar u$ for the special case of self-affine fractal surfaces
was studied in Ref. \cite{preparation} using $W(q,p)= \gamma P_{p}(q)$ in the expression for the elastic energy.

\begin{figure}
\includegraphics[width=0.45\textwidth,angle=0]{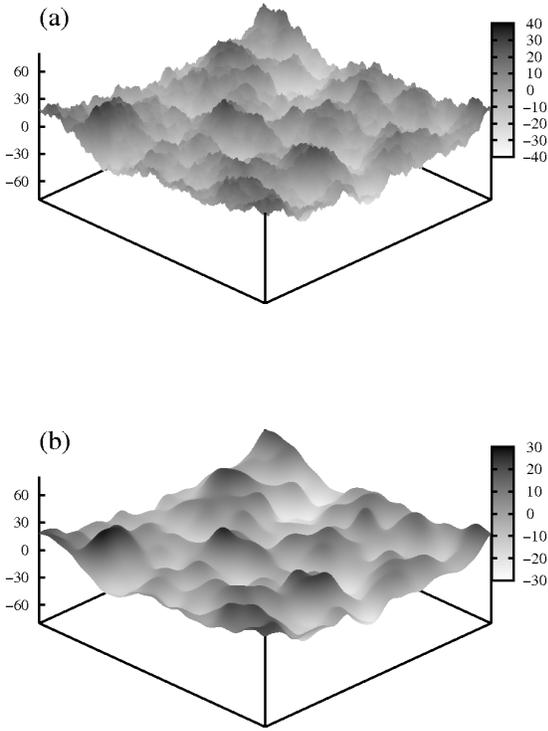}
\caption{\label{Topo.chunyan}
Surface height profile of a mathematically generated self affine fractal
surface ($104\times 104 \ {\rm nm}^2$ square surface area) with the root mean square roughness
$1 \ {\rm nm}$. 
(a) high magnification ($\zeta = 216$), (b) low magnification ($\zeta = 4$)}
\end{figure}

\begin{figure}
\includegraphics[width=0.45\textwidth,angle=0]{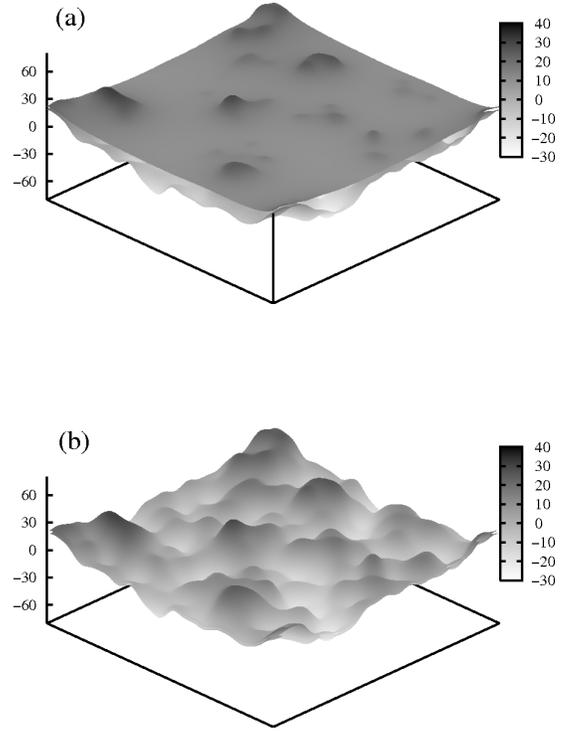}
\caption{\label{surfaces}
3D images of contacting regions with both substrate and block, at low magnification 
$\zeta=4$ under different load (a) $\sigma_{0}/E^*=0.0129$ and  $A/A_{0}=0.1089$, 
(b) $\sigma_{0}/E=0.26$ and $A/A_{0}=0.9527$. Note $E^*=E/(1-\nu^{2})$ is an effective modulus.}
\end{figure}

\begin{figure}
\includegraphics[width=0.45\textwidth,angle=0.0]{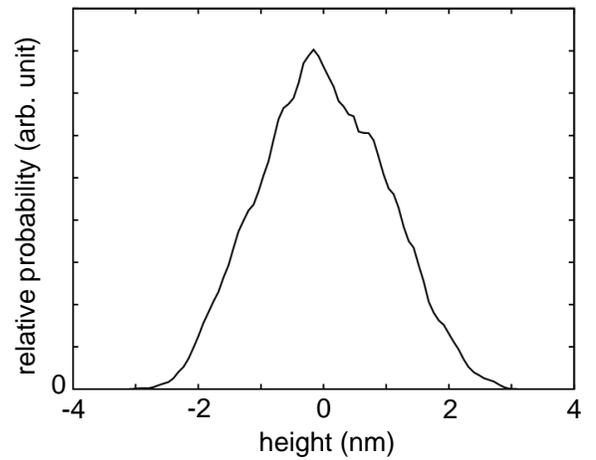}
\caption{\label{ChunyanPh}
The probability distribution $P_h$ of surface height $h$
for the mathematically generated surface shown in Fig. \ref{Topo.chunyan}.
For a square area $104 \times 104 \ {\rm  nm}^2$ with lattice constant
$a=1.605 \ {\rm \AA}$. 
}
\end{figure}

\begin{figure}
\includegraphics[width=0.45\textwidth,angle=0.0]{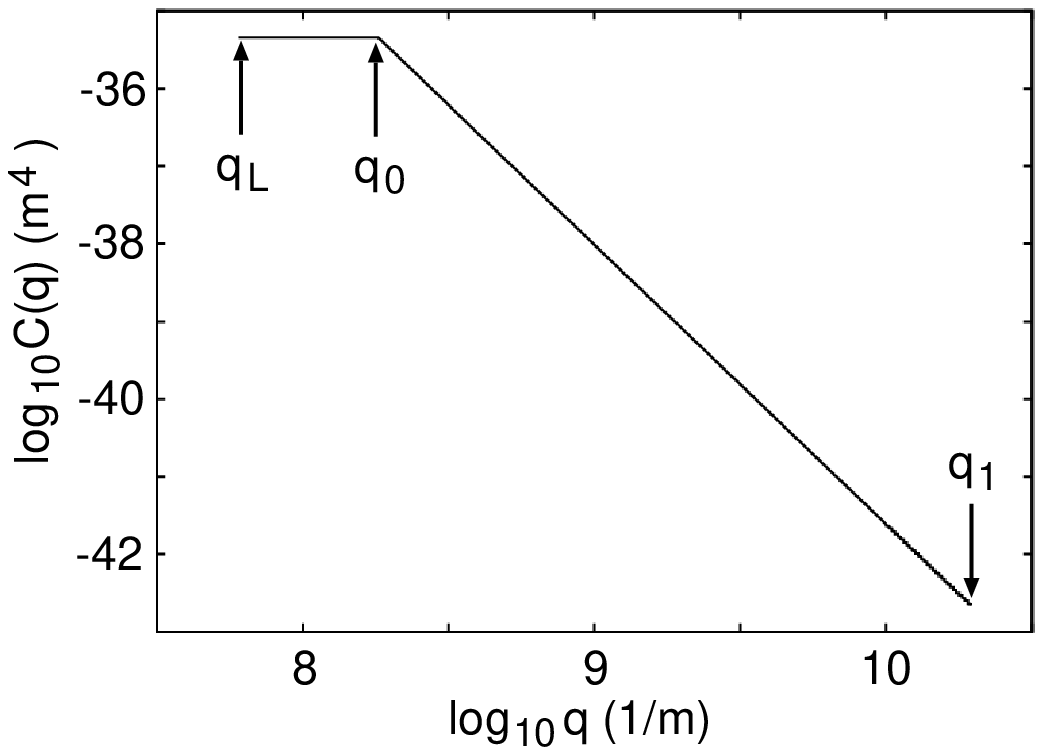}
\caption{\label{CqChunyan}
Surface roughness power spectra $C(q)$ of the mathematically generated substrate surface
studied in Ref. \cite{P3} and below. 
For a self affine fractal surface with the fractal dimension $D_{\rm f} = 2.2$ and
$q_L=6.04\times 10^7 \ {\rm m}^{-1}$, $q_0=1.81 \times 10^8  \ {\rm m}^{-1}$, 
and $q_1=3.9\times 10^{10} \ {\rm m}^{-1}$, 
and the root mean square roughness amplitude
$h_{\rm rms }= 1 \ {\rm nm}$.
}
\end{figure}

\vskip 0.3cm

{\bf 4. Molecular Dynamics}

We have performed Molecular Dynamics(MD) to study the contact area and the interfacial 
separation from small contact to full contact. We are interested in surfaces
with random roughness with wavelength components in some finite range 
$\lambda_0 > \lambda > \lambda_1$, where $\lambda_0$ is typically similar to 
(but smaller than) the lateral size of
the nominal contact area.
In order to accurately reproduce the contact mechanics between elastic 
blocks, it is necessary to consider solid block which extends at least a 
distance $\sim \lambda_0$ in the direction normal to the nominal contact area. 
(Note: the lower part of the system has been called substrate while the upper part
is called block.)
This leads to an enormous number of atoms or dynamical variables even for a small 
systems. In order to avoid this trouble we have developed a 
multiscale MD approach. This 
approach have been described in detail in Ref. \cite{Chunyan} and is only
summarized here. The system has lateral dimension $L_{x}=N_{x}a$ 
and $L_{y}=N_{y}a$, where $a$ is the lattice space of the block. 
Periodic boundary condition has been used in $xy$ plane.
For the block $N_{x}=N_{y}=400$, while the lattice space of the substrate
$b\approx a/\phi$, where $\phi=(1+\sqrt{5})/2$ is the golden mean, in order to avoid
the formation of commensurate structures at the interface. The mass of the block
atoms is 197 a.m.u. and the $a=2.6 \ \rm \AA$. 
The elastic modulus and Poisson ratio of the block is 
$E=77.2 \ {\rm GPa}$ and $\nu=0.42$, respectively. 

The atoms at the interface between the block and substrate interact with the repulsive potential
$U(r)=\epsilon \left( r_{0}/r \right)^{12}$,
where $r$ is the distance between a pair of atoms. 
We use $r_{0}=3.28 \ {\rm \AA}$ and $\epsilon=74.4 \ {\rm meV}$.
In the MD-model calculations there is no unique way of how to define the separation $\bar u$ between the
solid walls (see Ref. \cite{Chunyan} for a discussion of this point). 
We have used the same definition as in Ref. \cite{Chunyan} $\bar u=d-d_{c}$, where $d$ is the separation 
between the average $z$ coordinate of the bottom layer of the block atoms and the average plane of the 
substrate. $d_{c}$ is the critical atom-atom separation we use to define contact on atomic scale.
Thus, $u=0$ corresponds to the separation ${d_{c}=4.3615 \rm \AA}$ between 
planes through the center of mass of the interfacial atoms of the block and the substrate.

\begin{figure}
\includegraphics[width=0.45\textwidth,angle=0]{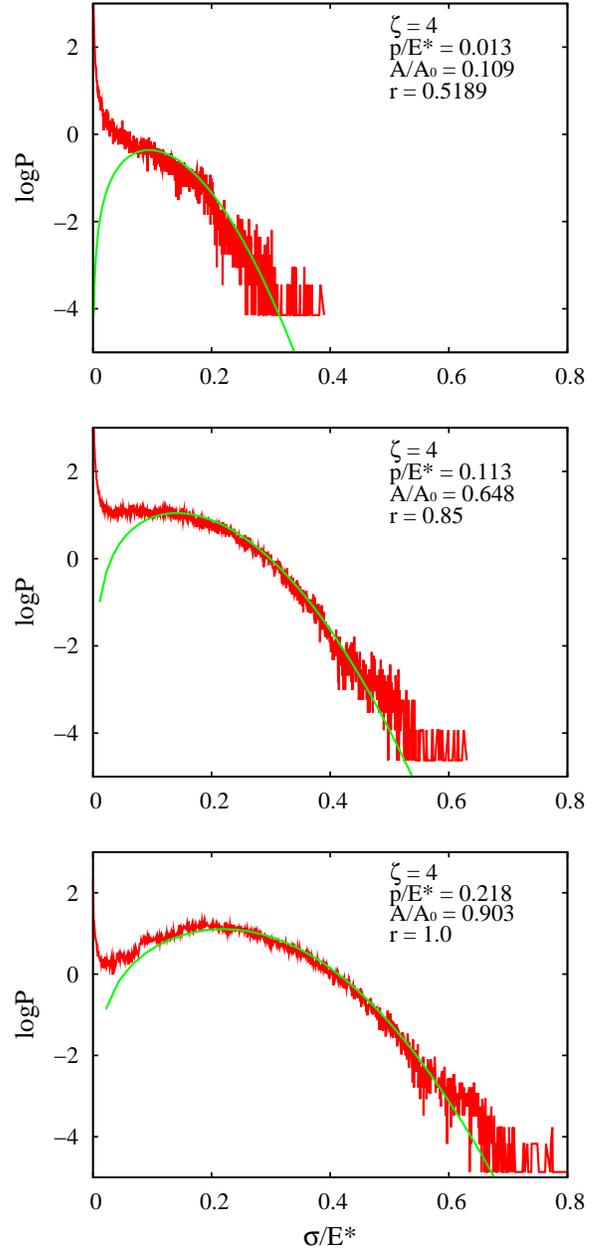}
\caption{\label{pressure_distribution}
The pressure distribution for $\zeta=4$ for three different nominal pressure.
The analytical theory has been used to fit the numerical pressure distribution 
as good as possible. Note $r=\tilde G/G$ (see Ref.\cite{MD.PRL}).}
\end{figure}

\begin{figure}
\includegraphics[width=0.45\textwidth,angle=0]{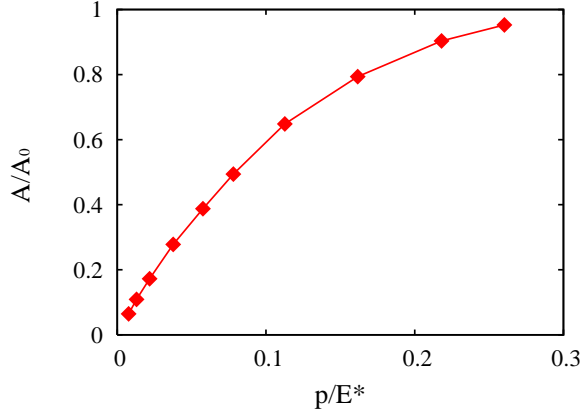}
\caption{\label{ratio}
Contact area ratio $A/A_{0}$ calculated from (21),
as a function of normalized pressure $p/E^*$.}
\end{figure}

\begin{figure}
\includegraphics[width=0.45\textwidth,angle=0]{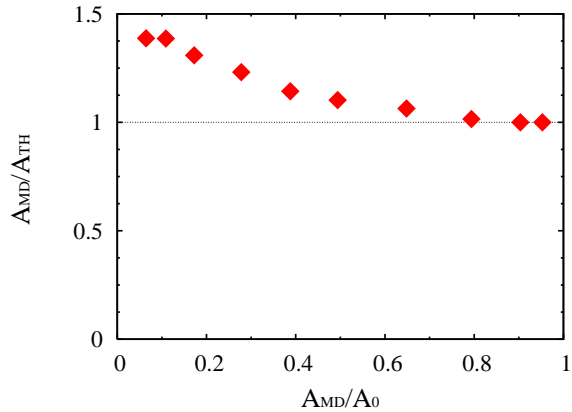}
\caption{\label{MD.TH.Com}
Projected contact area comparison between molecular dynamics simulations 
$A_{\rm MD}$ and continuum mechanics theory of Persson $A_{\rm TH}$. At small squeezing
pressure, $A_{\rm MD}$ is about $(30-38)\%$ bigger than the $A_{\rm TH}$. However, the 
difference decreases with increasing pressure.}
\end{figure}

\vskip 0.3cm

{\bf 5. Numerical results: Contact area}

From our molecular dynamics simulation we can calculate the interfacial stress distribution.
In order to obtain the contact area we follow the procedure outlined in Ref. \cite{Chunyan}
and fit the numerical results to the theoretically
predicted stress distribution
\begin{equation}
\label{eq20}
P(\sigma, \zeta) = {1\over 2 (\pi \tilde G)^{1/2}} 
\left (e^{-(\sigma -p)^2/4\tilde G}-e^{-(\sigma +p)^2/4\tilde G}\right ),
\end{equation}
where $\tilde G(p,\zeta)$ 
depends on the nominal squeezing pressure $p$ and the magnification $\zeta$
(but which is independent of $\sigma$),
and which we choose to get the best agreement with numerical data.
In Fig. \ref{pressure_distribution} we have shown
the good agreement between the numerical pressure
distribution and the analytical function (\ref{eq20})  
for $\zeta=4$ under three different nominal pressure.
When $\tilde G$ is known we can calculate 
the relative contact area using
$$ {A\over A_0} = \int_0^\infty  d\sigma \ P(\sigma,\zeta),
\eqno(21)$$
which is equivalent to using (4) with $G$ replaced by $\tilde G$.
In Fig. \ref{ratio} we show the result for $A/A_0$ 
as a function of normalized pressure $p/E^{*}$.

In Fig.{\ref{MD.TH.Com}} we have compared the contact area between MD 
simulations and continuum contact mechanics theory of Persson. Note that
the simulation for small load predict a contact area which is about $\sim 30\%$ larger 
than predicted by the theory. 
This is slightly larger than what has been found in earlier numerical simulations. Thus, the finite element
calculations of Hyun and Robbins\cite{Hyun2} and the Green's function molecular dynamics study of
Campana and M\"user\cite{Carlos} give a contact area
which is about $\sim 20 \%$ larger than that predicted by the Persson theory. 
Similarly, the study of H\"onig\cite{Hoen} gives a contact area which is about $\sim 25\%$ bigger than
the analytical one for small load. 
However, none of the computer simulations
can be considered as perfectly converged, and the difference between theory and fully converged
computer simulations 
may be smaller
than that indicated by the numbers given above. 
Thus, most of the numerical
studies reported use rather few grid points within the shortest wavelength roughness,
which results in an overestimation of the contact area \cite{Pei}. In our simulation for $\zeta=4$
we have many atoms within the shortest substrate roughness wavelength, 
but the surface roughness extends over less than one decade in length scale.  

Finally, we note that while the pressure distribution we obtain for low magnification ($\zeta=4$) 
is in a good agreement with Persson's theory, for the highest magnification 
($\zeta=216$) this is not the case because only one atom (or of order one atom)
occurs within the shortest wavelength roughness of the substrate (which is roughly
given by the substrate lattice constant). In the latter 
case we observe that the stress probability distribution for high normal stress
falls of roughly exponentially rather than like a Gaussian.
This has also been observed in earlier (non-converged) finite element calculations \cite{Hyun3}. 
It is clear that this limiting case cannot be described by the elastic
continuum model, which is the basis for all analytical contact mechanics theories.

\begin{figure}
\includegraphics[width=0.47\textwidth,angle=0]{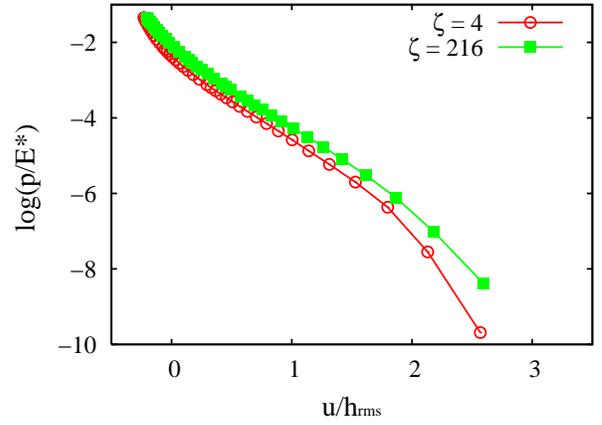}
\caption{\label{block}
An elastic block squeezed against a rigid rough substrate. The (natural) logarithm of the normalized 
average pressure $p/E^{*}$, as a function of the normalized 
separation between the average plane of the substrate and the average plane
of the lower surface of the block and the average plane of the lower surface 
of the block denoted by $\bar u/h_{rms}$. Adapted from \cite{MD.PRL}.}
\end{figure}

\begin{figure}
\includegraphics[width=0.45\textwidth,angle=0.0]{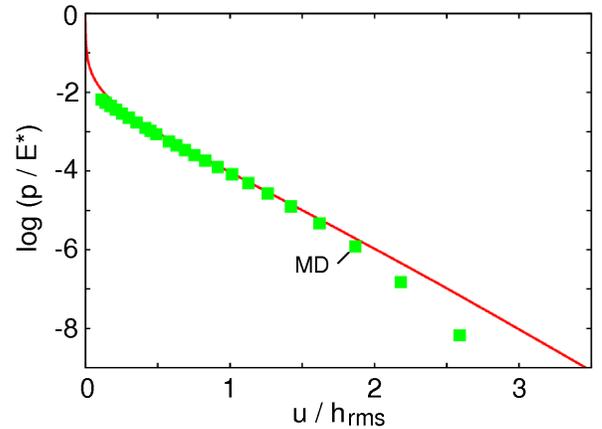}
\caption{\label{Chuny}
The relation between the (natural) logarithm of the
squeezing pressure $p$ (normalized by $E^{*}$) and the interfacial separation $\bar u$ (normalized
by the root-mean-square roughness amplitude $h_{\rm rms}$)
for an elastic solid squeezed against a rigid surface with the power spectra shown in 
Fig. \ref{Topo.chunyan}. In the calculation we have used
$\gamma = 0.42$. Adapted from \cite{MD.PRL}.
}
\end{figure}

\vskip 0.3cm

{\bf 6. Numerical results: Interfacial surface separation}

In Fig. \ref{block} we show the (natural) logarithm of the normalized 
average pressure $p/E^{*}$, as a function of the normalized 
separation $\bar u/h_{rms}$ between the average plane of the substrate and the average plane
of the lower surface of the block. We show results for the magnification $\zeta=4$ (open circles)
and $\zeta=216$ (solid squares). In this figure $\bar{u}=0$ corresponds to the separation $4.3615 \rm \AA$ between the
plane through the center of the atoms of the top layer of substrate atoms and bottom layer of
block atoms. Since the atoms interact with a long-range repulsive $\propto r^{-12}$ pair potential,
it is possible to squeeze the surfaces closer to each other than what corresponds to $\bar{u}=0$.
This explains why simulation data points occur also for $\bar{u} < 0$. The theory described in Sec.3
assumes a contact interaction potential so that $\bar{u} \ge 0$,
and can therefore not be compared with the MD 
simulations for very small (and negative) $\bar{u}$.  

In Fig. \ref{Chuny} we compare the MD results from Fig. \ref{block} (solid squares) with the theoretical
prediction calculated from (\ref{old8}) using the same surface roughness power spectra (and other
parameters) as in the MD-calculation. The theory is in good agreement with the numerical
data for $0.2 < \bar{u}/h_{rms} < 2$. For $\bar{u}/h_{rms} < 0.2$ the two curves differ because of the
reason discussed above, i.e., the ``soft'' potential used in the MD simulation allows
the block and substrate atoms to approach each other beyond $\bar{u}=0$, while in the analytical
theory a contact potential is assumed where the repulsive potential is
infinite for $\bar{u}<0$ and zero for $\bar{u}>0$. The difference between the theory and the
MD results for  $\bar{u}/h_{rms} > 2$ is due to a finite size effect. That is, since the MD calculations use a 
very small system, the highest asperities are only $\sim 3 h_{rms}$ above the average plane (see the height distribution in
Fig. \ref{ChunyanPh}), and for large $\bar{u}$ very few contact spots will occur, and in particular for $\bar{u} > 3 h_{rms}$
no contact occurs and $p$ must vanish. In the analytical theory, the system is assumed to be infinite
large. So that even for a Gaussian distribution of asperity height, there will always be (infinitely) many 
infinitely high asperities and contact will occur at arbitrary large separation $\bar{u}$. The
asymptotic relation $\bar u\propto {\rm log} p$ will hold for arbitrarily large $\bar{u}$ (or small squeezing pressures
$p$).

\begin{figure}
\includegraphics[width=0.35\textwidth,angle=-90.0]{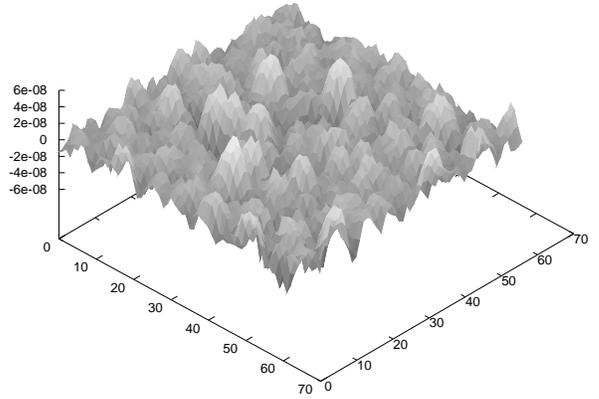}
\caption{\label{topo}
Surface height profile of a polymer surface, with the root mean square roughness
$14 \ {\rm nm}$, measured over a $10\times 10 \ {\rm \mu m}^2$ square surface area. 
}
\end{figure}

\begin{figure}
\includegraphics[width=0.45\textwidth,angle=0.0]{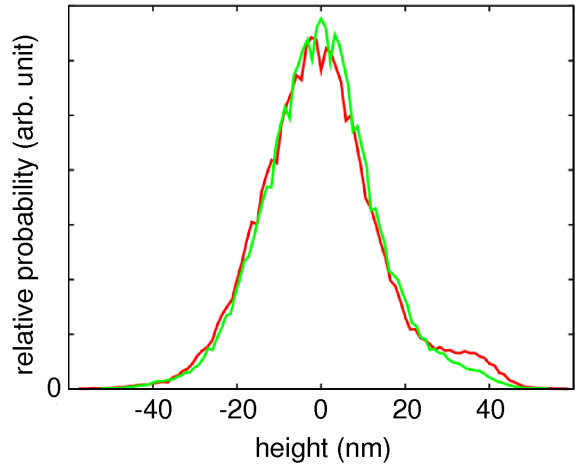}
\caption{\label{Ph}
The probability distribution $P_h$ of surface height $h$
for the polymer film in Fig. \ref{topo}.
The two curves correspond to height profile data measured at two different 
square surface
areas, each $10 \times 10 \ {\rm \mu m}^2$. 
}
\end{figure}

\begin{figure}
\includegraphics[width=0.45\textwidth,angle=0.0]{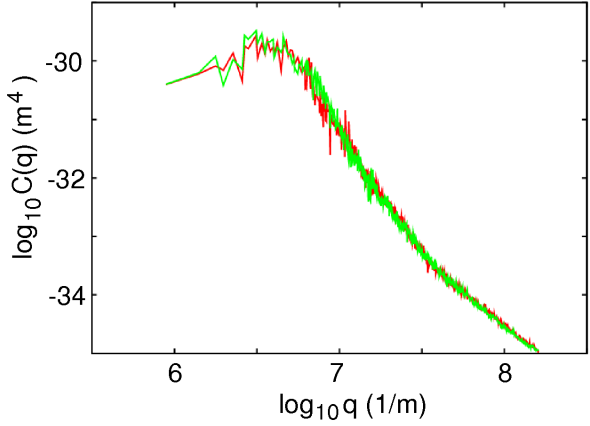}
\caption{\label{Cq}
Surface roughness power spectra $C(q)$ of the polymer film in Fig. \ref{topo}.
The two curves correspond to height profile data measured at two different square surface
areas, each $10 \times 10 \ {\rm \mu m}^2$. The height probability distribution for the same surface areas
is shown in Fig. \ref{Ph}. 
}
\end{figure}

\begin{figure}
\includegraphics[width=0.45\textwidth,angle=0.0]{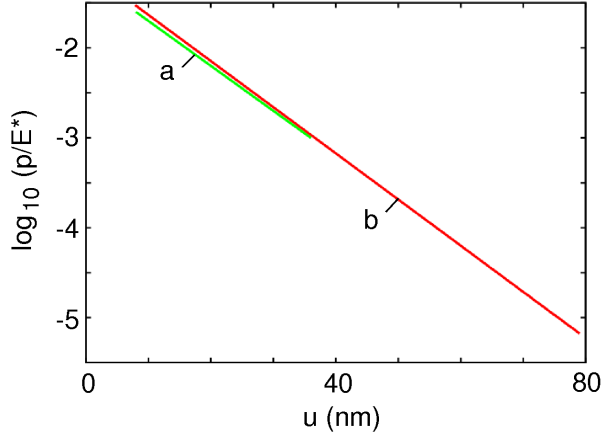}
\caption{\label{Rob1}
The relation between the logarithm (with 10 as basis) of the
squeezing pressure $p$ (normalized by $E^*$) and the interfacial separation $\bar{u}$ (in nm)
for an elastic solid squeezed against a rigid surface with the power spectra given by the sum of the
two power spectra shown in 
Fig. \ref{Cq}. The line {\bf a} is the result
of the Finite Element calculation of Pei et al, while the line {\bf b} is the
prediction of the theory. 
In the calculation we have used
$\gamma = 0.38$.
}
\end{figure}

\begin{figure}
\includegraphics[width=0.45\textwidth,angle=0.0]{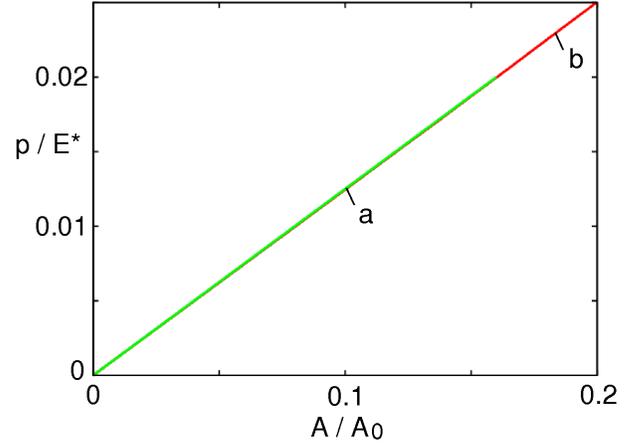}
\caption{\label{Rob2}
The relation between the relative contact area, $A/A_0$, and the normalized 
squeezing pressure, $p/E^*$, 
for an elastic solid squeezed against a rigid surface with the power spectra 
given by the sum of the two power spectra shown in
Fig. \ref{Cq}. The line {\bf a} is the result
of the Finite Element calculation of Pei et al, while the line {\bf b} is the
prediction of the theory where $A/A_0$ is scaled by a factor of $1.29$.
}
\end{figure}

\vskip 0.3cm

{\bf 7. Contact mechanics for a measured surface}

Pei et al\cite{Pei} have performed a Finite Element Method (FEM)
computer simulation of the contact mechanics
for a polymer surface, using the measured
surface topography\cite{Benz} as input, squeezed against a flat surface. 
Here we would like to compare the FEM results with the analytical results of Persson.

In Fig. \ref{topo} we show the 
surface height profile of a polymer surface studied in Ref. \cite{Benz} with the root mean square roughness
$14 \ {\rm nm}$. The probability distribution of surface height $P_h$, for two different 
square $10\times 10 \ {\rm \mu m}^2$ surface areas on the polymer film, has been shown in Fig. \ref{Ph}.
These two surface areas were used in the FEM calculation by Pei et al as the upper and lower
surface. It is remarkable that in spite of the fact that the height distributions are not perfect
Gaussian (in particular one surface exhibits a ``bump'' in the height distribution on the
outer side of the height profile), Pei et al obtained a nearly perfect linear relation
between ${\rm log} p$ and $\bar{u}$. This result indicates that even in the present case
the area of real contact is 
proportional to the load (see below) 
and the statistical properties of the contact regions does not change with the load for small load.  
The surface roughness power spectra of the two surfaces are shown in Fig. \ref{Cq}, and are almost identical
in spite of the difference which occurs in the height distribution $P_h$.

Fig. \ref{Rob1}
shows the relation between the logarithm (with 10 as basis) of the
squeezing pressure $p$ (normalized by $E^*$) and the interfacial separation $\bar{u}$ (in nm)
for an elastic solid squeezed against a rigid surface with the power spectra given by the sum of
the two power spectra shown in 
Fig. \ref{Cq}. The line {\bf a} is the result
of the FEM calculation of Pei et al, and shows that 
for large separation $p\propto {\rm exp} (-\bar u/\gamma h_{\rm rms})$ with $\gamma \approx 0.4$,
which is consistent with our analytical results 
(see line {\bf b} in Fig. \ref{Rob1}).

Fig. \ref{Rob2}
shows the relation between the relative contact area, $A/A_0$, and the normalized 
squeezing pressure, $p/E^*$, 
for an elastic solid squeezed against a rigid surface with the power spectra, given by the sum 
of two power spectra shown in 
Fig. \ref{Cq}. The line {\bf a} is the result
of the FEM calculation of Pei et al, while the line {\bf b} is the
prediction of the theory where $A/A_0$ is scaled by a factor of $1.29$. 

The good agreement, between the FEM calculations and the analytical theory found above,
indicates that the contact mechanics results are robust and not very sensitive to many details
such as the assumption of perfectly random surfaces, which is unlikely to be exactly obeyed
for the polymer surfaces, the surface topography of which was used in the FEM calculations. 
(In order to address to what extent a measured surface is randomly
rough, one would need to calculate higher order correlation functions, and show that
odd order functions in the height coordinate vanish (or are very small), 
and that even order height correlation
functions can be decomposed into a sum of pair correlation functions. 
We are aware of no such
study for ``real'' measured surface profiles.)

Finally we note that the observation of an effective exponential repulsion has important 
implications for tribology, colloid science,
powder technology, and materials science\cite{Benz}. 
For example, the density or volume of granular materials has long been known to have a
logarithmic dependence on the externally applied isotropic pressure or stress, as found, for example,
in the compression stage during processing of ceramic materials\cite{Stanley}. Recent work on the confinement of
nanoparticles has also indicated an exponential force upon compression\cite{Alig},
suggesting that this relationship could be prevalent among quite different 
types of heterogeneous surfaces. 

\begin{figure}
\includegraphics[width=0.35\textwidth,angle=0.0]{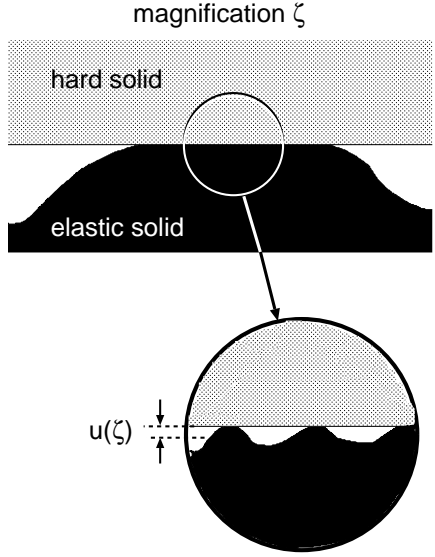}
\caption{\label{asperity.mag}
An asperity contact region observed at the magnification $\zeta$. It appears that
complete contact occur in the asperity contact region, but upon increasing the magnification
it is observed that the solids are separated by the average distance $\bar{u}(\zeta)$.
}
\end{figure}

\begin{figure}
\includegraphics[width=0.45\textwidth,angle=0.0]{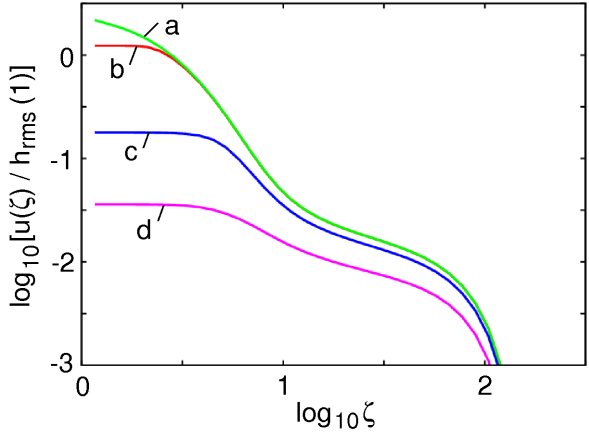}
\caption{\label{log.zeta.u}
The dependence on the magnification,
of the 
the average distance $\bar{u}(\zeta)$ (in units of the root-mean-square
roughness amplitude $h_{\rm rms}(1)$ of the whole surface)
between the surfaces in  
an asperity contact region observed at the magnification $\zeta$. 
For the polymer surface with the power spectra shown in Fig. \ref{Cq}.
We have assumed the effective elastic modulus $E^* = 2 \ {\rm GPa}$ and the
nominal squeezing pressure {\bf a}: $p_0=1$, {\bf b}: $10$, {\bf c}: 
$100$ and {\bf d}: $200 \ {\rm MPa}$.
}
\end{figure}

\begin{figure}
\includegraphics[width=0.45\textwidth,angle=0.0]{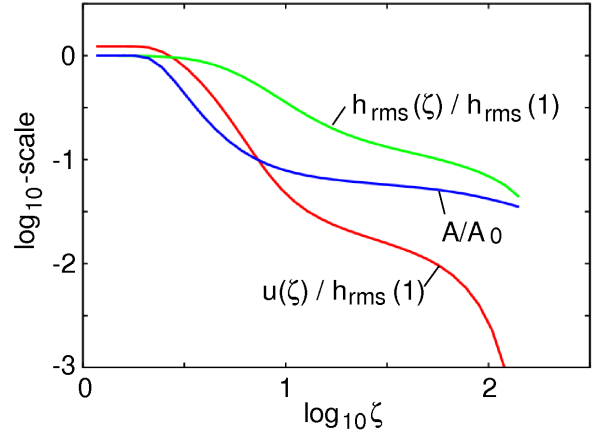}
\caption{\label{uzeta}
The dependence on the magnification,
of the relative contact area $A(\zeta)/A_0$,
the root-mean-square roughness amplitude $h_{\rm rms}(\zeta)$ (in units of the root-mean-square
roughness amplitude $h_{\rm rms}(1)$ of the whole surface) and the average distance $\bar{u}(\zeta)$
between the surfaces in an 
an asperity contact region observed at the magnification $\zeta$. 
All quantities are shown on a logarithmic (with 10 as basis) scale.
For the polymer surface with the power spectra shown in Fig. \ref{Cq}.
We have assumed the effective elastic modulus $E^* = 2 \ {\rm GPa}$ and the
nominal squeezing pressure $p_0=10 \ {\rm MPa}$.
}
\end{figure}

\vskip 0.3cm

{\bf 8. Variation of the average surface separation $\bar{u}(\zeta)$ with the 
magnification $\zeta$} 

The theory presented above can be easily generalized in various ways.
Thus, it is possible to include the adhesional interaction\cite{Full,Johnson1}.
In this case the work done by the external pressure $p$ will be the sum of the stored (asperity
induced) elastic energy plus the (negative) adhesional energy, i.e., the right hand side of
(\ref{equal}) will now be $U_{\rm el}+U_{\rm ad}$. 
The theory can also be 
applied to study how the spacing $\bar{u}(\zeta)$ depends on the magnification. Here 
$\bar{u}(\zeta)$ is the (average) spacing between the solids in the apparent contact areas
observed at the magnification $\zeta$. 

Fig. \ref{asperity.mag} shows an asperity contact region 
at the magnification $\zeta$. It appears that
complete contact occurs in the asperity contact region, but upon increasing the magnification
it is observed that the solids are separated by the average distance $\bar{u}(\zeta)$.
The information about how $\bar{u}(\zeta)$ depends on the magnification is crucial for many
important applications, e.g., sealing (see below).

We study the contact between the solids at increasing magnification. In an apparent contact area
observed at the magnification $\zeta$, the substrate has the root mean square roughness amplitude
$$h_{\rm rms}^2 (\zeta) = 2\pi \int_{\zeta q_0}^{q_1} dq \ q C(q)$$
The separation between the surfaces $\bar{u}(\zeta)$ is given by (14) but with the lower
integration limit given by $q_0\zeta$ instead of $q_0$ and the lower pressure integration limit given by 
$p(\zeta) =p_0 A_0/A(\zeta)$: 
$$\bar{u}(\zeta ) = \surd \pi \int_{\zeta q_0}^{q_1} dq \ q^2C(q) 
w(q)$$
$$\times \int_{p(\zeta)}^\infty dp' 
 \ {1 \over p'} 
\left [\gamma+3(1-\gamma)P^2(q,p',\zeta)\right ] e^{-[w(q,\zeta) p'/E^*]^2}\eqno(23)$$
where
$$w(q,\zeta)=\left (\pi \int_{\zeta q_0}^q dq' \ q'^3 C(q') \right )^{-1/2}\eqno(24)$$
and where $P(q,p',\zeta)$ is given by (12) with $s=w(q,\zeta)/E^*$.
When we study the apparent contact area at increasing magnification, the contact pressure
$p(\zeta)$ will increase and the surface roughness amplitude $h_{\rm rms} (\zeta)$ will decrease. Thus,
the average separation $\bar{u}(\zeta)$, between the surfaces in the (apparent) contact regions observed at
the magnification $\zeta$, will decrease with increasing magnification.

In Fig. \ref{log.zeta.u} we show, 
for the polymer surface with the power spectra shown in Fig. \ref{Cq}, how 
the logarithm of the average distance $\bar{u}(\zeta)$ (in units of the root-mean-square
roughness amplitude $h_{\rm rms}(1)$ of the whole surface), depends on the magnification $\zeta$. 
We have assumed the effective elastic modulus $E^* = 2 \ {\rm GPa}$ and the
nominal squeezing pressures {\bf a}: $p_0=1$, {\bf b}: $10$, {\bf c}: 
$100$ and {\bf d}: $200 \ {\rm MPa}$. Note that for $\zeta > 2$ the two lowest
squeezing pressures give virtually identical separation between the surfaces in the (apparent) asperity
contact regions, in spite of the fact that the pressure for curve {\bf b} is 10 times higher than
for curve {\bf a}. This just reflects the fact stated earlier that for low squeezing pressure
the area of (apparent) contact $A$ varies linearly with
the squeezing force $pA_0$, and the interfacial stress distribution, and the 
size-distribution of contact spots, and the interfacial separation $\bar{u}(\zeta)$, 
are independent of the squeezing pressure\cite{Arch,PSSR}. 
That is, with increasing $p$,
existing contact areas grow and new contact areas form in such a way that in the thermodynamic limit
(infinite-sized system) the quantities referred to above remain unchanged.

In Fig. \ref{uzeta} we again show how the average distance $\bar{u}(\zeta)$ depends on the magnification $\zeta$.
We also show the $\zeta$-dependence of
the relative contact area $A(\zeta)/A_0$, and
the root-mean-square roughness amplitude $h_{\rm rms}(\zeta)$ (in units of the root-mean-square
roughness amplitude $h_{\rm rms}(1)$ of the whole surface).
All quantities are shown on a logarithmic (with 10 as basis) scale, and the results are for
the polymer surface with the power spectra shown in Fig. \ref{Cq}.
We have assumed the effective elastic modulus $E^* = 2 \ {\rm GPa}$ and the
nominal squeezing pressure $p_0=10 \ {\rm MPa}$.
Note that for $\zeta < 2$,  $\bar{u}(\zeta)/h_{\rm rms}(1) \approx 1.23$ i.e., at the lowest magnification
the upper surface is ``riding'' on top of the largest substrate asperities. The separation
between the solids in the asperity contact regions rapidly drop with increasing magnification
(corresponding to smaller and smaller asperity contact regions), and already at the magnification $\zeta=10$
the separation is only $\sim 3 \%$ of the average separation which occurs at the lowest magnification.
This is, of course, mainly due to the strong increase in the local pressure 
(as manifested in the decreased contact area, $A(10)/A_0 \approx 0.1$) in the asperity contact regions
at high magnification but also due to the reduction in the effective roughness detected over short distances
(the largest contribution to $h_{\rm rms} (1)$ comes from the longest wavelength roughness components).

\begin{figure}
\includegraphics[width=0.45\textwidth,angle=0.0]{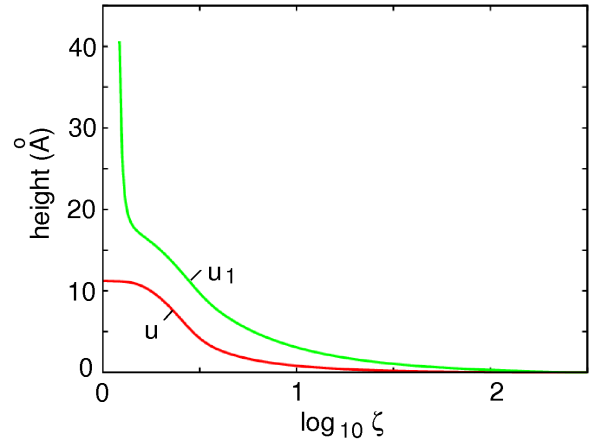}
\caption{\label{height.u.u1}
The interfacial separation $\bar{u}(\zeta)$ and ${u}_1(\zeta)$ as a function of
the magnification $\zeta$ for the surface shown in Fig. \ref{Topo.chunyan} and for the
squeezing pressure $1.3 \ {\rm GPa}$.}
\end{figure}

\begin{figure}
\includegraphics[width=0.45\textwidth,angle=0.0]{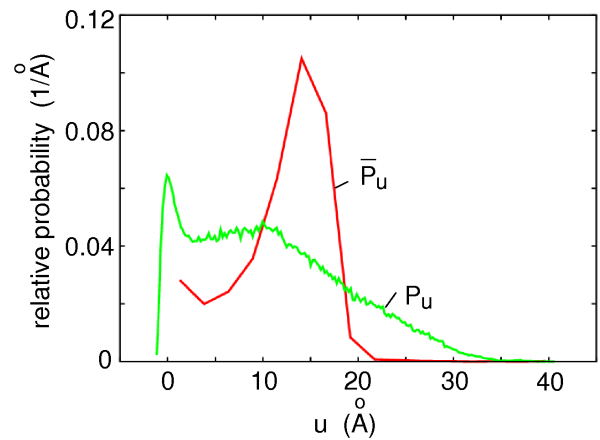}
\caption{\label{both.P.barP}
The probability distributions $P_u$ and $\bar P_u$ are defined in the text.
For the surface shown in Fig. \ref{Topo.chunyan} and for the
squeezing pressure $1.3 \ {\rm GPa}$. The distribution $\bar P_u$ has a delta function
at $u=0$ with the weight $A(\zeta_1)/A_0 \approx 0.047$.}
\end{figure}

Let ${u}_1(\zeta)$ be the (average) height separating the surfaces which appear to come into
contact when the magnification decreases from $\zeta$ to $\zeta-\Delta \zeta$, where $\Delta \zeta$
is a small (infinitesimal) change in the magnification. The empty volume between the surfaces which
appears to be in contact at the magnification $\zeta-\Delta \zeta$ is given by 
$\bar{u}(\zeta-\Delta \zeta) A(\zeta -\Delta \zeta)$. But this volume must be the sum of the volume
$ \bar{u}(\zeta) A(\zeta)$ between the surfaces which appears to be in contact at the magnification $\zeta$, plus
the additional volume ${u}_1(\zeta)[A(\zeta-\Delta \zeta)-A(\zeta)]$ which results from the increase in the
apparent contact area as the magnification decrease 
from $\zeta$ to $\zeta-\Delta \zeta$:
$$\bar{u}(\zeta-\Delta \zeta) A(\zeta -\Delta \zeta) = \bar{u}(\zeta) A(\zeta) 
+{u}_1(\zeta)[A(\zeta-\Delta \zeta)-A(\zeta)]$$
or
$${u}_1(\zeta) = \bar{u}(\zeta)+\bar u'(\zeta) A(\zeta)/A'(\zeta)\eqno(25)$$
In Fig. \ref{height.u.u1} we show 
the interfacial separation $\bar{u}(\zeta)$ and ${u}_1(\zeta)$ as a function of
the magnification $\zeta$ for the surface shown in Fig. \ref{Topo.chunyan} and for the
squeezing pressure $1.3 \ {\rm GPa}$. Note that for small magnification $u_1$ increases
rapidly. This can be understood as follows:
The theory is for an infinite system. For an infinite system,
even for a Gaussian distribution of surface height, there will always be some
infinitely high asperities and some infinitely deep valleys. Thus during contact there
will always be some regions where the surface separation is arbitrarily high.
Of course, the fraction of the surface where $u_1$ is large is extremely small. 
The reason is that the variation of the 
(apparent) contact area with the magnification is negligible in the region where
$u_1$ start to grow fast (see Fig. \ref{height.u.u1}). 
Thus, the strong increase in $u_1$ for small magnification is of no practical importance--it
is a purely academic result. 

The quantity ${u}_1(\zeta)$ is very important in the context of the leakage 
through rubber sealing: Let us study the interface between the rubber and the substrate
as the magnification increases. At low magnification it appears as if the solids makes perfect 
contact at the interface. As we increase the magnification
non-contact area becomes visible.
At large enough magnification, say $\zeta = \zeta^*$, the non-contact area will percolate\cite{P3}. 
A first rough estimate of the gas (or fluid) leakage is obtained by assuming that the gas
flow through a pipe or pore with width and length $\lambda \approx L/\zeta^*$ 
(where $L$ is the linear size of the 
sealing contact area) and height ${u}_1(\zeta^*)$, and that the
whole pressure drop $\Delta p = p_1-p_0$ (where $p_1$ and $p_0$ is the 
pressure to the left and right of the
sealing) occurs over the pore. Thus for an incompressible
fluid, the mass-flow per unit time through the interfacial pore
will be $\dot Q \approx \rho u_1^3(\zeta^*) \Delta p/12 \eta$ (where $\eta $ is the viscosity). 
We will analyze this problem in greater detail 
elsewhere\cite{Perssontbp}.

Finally, let us consider the distribution of interfacial separations\cite{MD.PRL} 
$$P_u = \langle \delta (u-u({\bf x}))\rangle \eqno(26)$$
where $\langle .. \rangle$ stands for ensemble average (which often is equivalent to 
average over the surface area).
We also define another distribution $\bar P(u)$ of interfacial separations which differ from (26)
by using instead of $u({\bf x})$ another function which involves some average over the spatial
coordinate and defined as follows:
The probability to find the surface separation $u < u_1(\zeta)$ is
$$\Psi (u) = A(\zeta)/A_0 = P(\zeta)$$
where $\zeta=\zeta(u)$ is the solution to $u=u_1(\zeta)$
[note: $u_1(\zeta)$ is a monotonically decreasing function of $\zeta$
so there exists a unique solution $\zeta = \zeta(u)$ to $u=u_1(\zeta)$].
The probability distribution
$$\bar P_u = {d\Psi (u) \over du} = {P'(\zeta) \over u'_1(\zeta)}|_{\zeta=\zeta(u)}$$
We can also write
$$\bar P_u = -\int d\zeta {A'(\zeta) \over A_0} \delta \left [ u-u_1(\zeta)\right ]\eqno(27)$$
It is convenient to change integration variable to $\mu$ defined by
$\zeta = {\rm exp}({-\mu})$ and consider $A$ and $u_1$ as function of $\mu$. This gives
$$\bar P_u = -\int d\mu {A'(\mu) \over A_0} \delta \left [ u-u_1(\mu)\right ]\eqno(28)$$
In Fig. \ref{both.P.barP} we show $P_u$ and $\bar P_u$
for the surface shown in Fig. \ref{Topo.chunyan}, and for the
squeezing pressure $1.3 \ {\rm GPa}$. The result for $P_u$ was obtained from the MD-simulations,
while $\bar P_u$ was obtained from the analytical theory presented above. 
As expected the distribution $\bar P_u$ is more narrow
than $P_u$ [since it involves $u_1$ which is already an average of $u({\bf x})$],
but it is easy to show that average of $u$ is the same for both $P$ and $\bar P$ and equal to
the average separation between the surfaces. Thus
$$\int du \ u \bar P_u = -\int d\zeta \ {A'(\zeta )\over A_0} u_1(\zeta)$$
Substituting (25) in this equation gives
$$\int du \ u \bar P_u = -{1\over A_0} \int_{\zeta_0}^{\zeta_1} d\zeta \ {d \over d \zeta} [A(\zeta) \bar u(\zeta)] 
= \bar u(\zeta_0)$$
where we have used that $\bar u(\zeta_1)=0$ and that $A(\zeta_0)=A_0$ (note: $\zeta_0$ is the lowest magnification
usually chosen as unity). 
Since $\bar u(\zeta_0)$ equals the average separation
between the surfaces we have proved our statement. Note that $\bar P_u$ has a delta function at
$u=0$ with the weight $A(\zeta_1)/A_0$. Using this fact it is easy to show that distribution
$\bar P_u$ has the zero order moment equal to unity:
$$\int du \ \bar P_u = 1$$
Thus, the zero order and first order moments of $P_u$ and $\bar P_u$ are the same, but higher order
moments will differ.

\begin{figure}
\includegraphics[width=0.4\textwidth,angle=0.0]{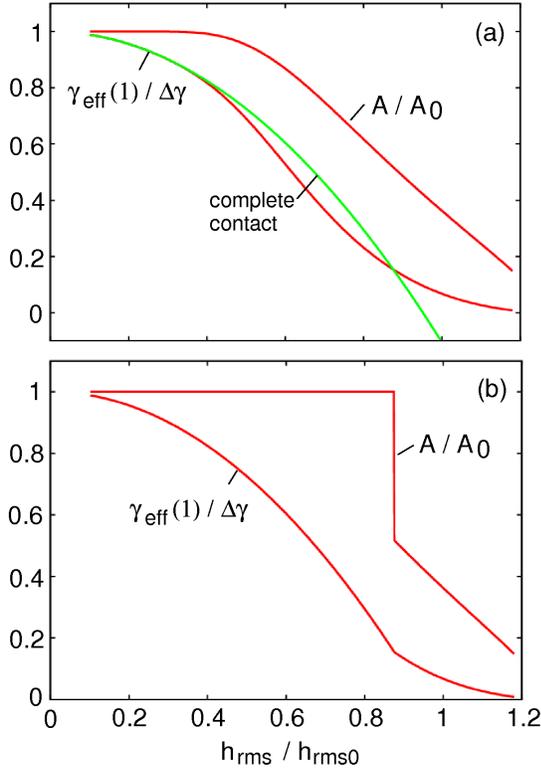}
\caption{\label{adhesionfig}
The effective interfacial binding energy and the area of real contact
as a function of the normalized root-mean-square roughness $h_{\rm rms}/h_{\rm rms0}$. The case
$h_{\rm rms}=h_{\rm rms0}$ corresponds to the power spectra obtained from the measured height profile
for the surface ${\bf 7}$ (with $h_{\rm rms0} \approx 0.2 \ {\rm \mu m}$)
studied in Ref. \cite{Pere} (see also Fig. \ref{Rubber.ball.Figure.3}). In the calculation
we have used the measured (low-frequency) elastic modulus ($E\approx 5 \ {\rm MPa}$) 
and the measured (for flat surfaces) interfacial binding energy (per unit area) 
($\Delta \gamma \approx 0.1 \ {\rm J/m^2}$). In (a) we show the calculated $\gamma_{\rm eff} (1)$
(red curve) and the atomic contact area  $A(\zeta_1)/A_0$ observed at the highest magnification.
The green curve gives $\gamma_{\rm eff} (1)$ under the assumption that complete contact 
occurs at the interface. When the system is in thermal equilibrium it will be in the state where
$\gamma_{\rm eff}$ is maximal. In (b) we show the thermal equilibrium interfacial binding energy
and the corresponding contact area. Note that the system flips abruptly from the complete contact
to partial (about $50 \%$) contact at $h_{\rm rms} \approx 0.9 h_{\rm rms0}$.  
}
\end{figure}

\begin{figure}
\includegraphics[width=0.45\textwidth,angle=0.0]{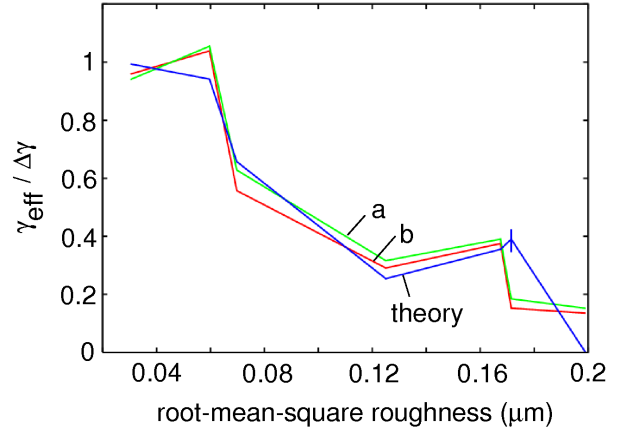}
\caption{\label{Rubber.ball.Figure.3}
The normalized effective interfacial binding energy $\gamma_{eff}/\Delta\gamma$ as a function of the root-mean-square roughness for 7 differently
prepared surfaces. Blue curve: theory, assuming perfect (atomic) contact at the interface; 
green curve: {\bf a}; and red curve: {\bf b}.
The experimental data for the pull-off velocity $0.2$ and $2 \ {\rm \mu m/s}$.
From ref. \cite{Pere}}.
\end{figure}

\vskip 0.3cm

{\bf 9. Elastic energy and adhesion}

In Sec. 3 we have shown that from the
dependence of the surface separation $u$ on the squeezing pressure $p$, deduced from, 
e.g., experiments or from MD or FEM calculations,
one can obtain the elastic energy
$U_{\rm el}$ stored in the asperity contact regions. 
Thus, in Sec. 3 we have presented
an expression for $U_{\rm el}$ which results in nearly the same relation between $u$ and $p$ as observed
in the numerical MD and FEM studies.
This is an important result as $U_{\rm el}$ is relevant for 
many important applications, e.g., adhesion between elastic solids 
with randomly rough surfaces\cite{P1,JPCM2006,Carbone}.

We consider the adhesive contact between two elastic solids with randomly rough surfaces.
Assume that the surface roughness power spectra has a long-distance roll-off wavelength
$\lambda_0$ (corresponding to the roll-off wave vector $q_0 = 2 \pi /\lambda_0$) which is much
smaller than the diameter of the nominal contact area. In this case we can take into account
the influence of the surface roughness on the (adhesive) contact mechanics by 
using an effective interfacial binding energy\cite{P1}
$$\gamma_{\rm eff}(\zeta) A^*(\zeta) = \Delta \gamma A^*(\zeta_1)-U_{\rm el}(\zeta),$$
where $U_{\rm el}(\zeta)$ is the elastic energy stored in the asperity contact region
as a result of the asperities which cannot be observed at the magnification $\zeta$ (i.e.,
due to the surface roughness with wavelength shorter than $\lambda_0/\zeta$).
$A^*(\zeta)$ is the contact area when the surface is studied at the magnification $\zeta$, which
in general is larger than the projected contact area $A(\zeta)$.
The interfacial binding energy per unit surface area for the contact between 
two perfectly flat surfaces of the two solids is denoted by $\Delta \gamma$, and $A^*(\zeta_1)$
is the contact area observed at the highest (atomic) magnification $\zeta_1$. Using (9) we can
write\cite{P1}
$$\gamma_{\rm eff}(\zeta) = \Delta \gamma {A^*(\zeta_1)\over A^*(\zeta)} -
E^* {\pi \over 2} {A_0\over A^*(\zeta)} \int_{q_0\zeta}^{q_1} dq \ q^2W(q,p)C(q)$$
The {\it macroscopic} effective interfacial binding energy $\gamma_{\rm eff}(1)$ determines the 
macroscopic contact mechanics and the pull-off force. For example, for a rubber ball 
(radius $R$) in contact
with a nominal flat substrate, the pull-off force is given by the JKR formula
$F_{\rm pull-off} = (3\pi /2)R \gamma_{\rm eff}$. Thus, 
the surface roughness enters in the expression for $\gamma_{\rm eff}(1)$, but the surfaces
can be considered as perfectly smooth when solving the macroscopic contact mechanics problem.

In Ref. \cite{P1} one of us studied $\gamma_{\rm eff}(\zeta)$ using the 
approximation $W(q,p)=P_{p}(q)$ when calculating the elastic energy.
Here we will present results using the improved expression 
for $U_{\rm el}$ with $W(q,p)$ given by (10).
In Fig. \ref{adhesionfig} we show
the effective binding energy and the area of real contact
as a function of the normalized root-mean-square roughness $h_{\rm rms}/h_{\rm rms0}$. The case
$h_{\rm rms}=h_{\rm rms0}$ corresponds to the power spectra obtained from the measured height profile
for the surface ${\bf 7}$ (with $h_{\rm rms0} \approx 0.2 \ {\rm \mu m}$)
studied in Ref. \cite{Pere} (see also Fig. \ref{Rubber.ball.Figure.3}). 
The power spectra used in Fig. \ref{adhesionfig} have been obtained
by scaling the power spectrum of surface {\bf 7} with the 
factor $(h_{\rm rms}/h_{\rm rms0})^2$.
In the calculation of $\gamma_{\rm eff}$
we have used the measured (low-frequency) elastic modulus ($E\approx 5 \ {\rm MPa}$) 
and the measured (for flat surfaces) interfacial binding energy (per unit area) 
($\Delta \gamma \approx 0.1 \ {\rm J/m^2}$). 
In Fig. \ref{adhesionfig}(a) we show the calculated $\gamma_{\rm eff} (1)$
(red curve) and the atomic contact area  $A(\zeta_1)/A_0$ observed at the highest magnification.
The green curve gives $\gamma_{\rm eff} (1)$ under the assumption that complete contact 
occurs at the interface. When the system is in thermal equilibrium it will be in the state where
the interfacial binding energy
$\gamma_{\rm eff}$ is maximal. In Fig. \ref{adhesionfig} (b) we  
show the thermal equilibrium interfacial binding energy
and the corresponding contact area. Note that the system flips abruptly from the complete contact
to partial (about $50 \%$) contact at $h_{\rm rms} \approx 0.9 h_{\rm rms0}$.  
Very similar abrupt transitions have been found in exact solution for a cosines roughness
profile, see e.g., Ref. \cite{Zil}. We also note that while Fig. \ref{adhesionfig}(b) shows the minimum free 
energy state, hysteresis may occur in practical situations (see Ref. \cite{Zil}).
Finally, note that the contact area $A(\zeta_1)$ is finite also when $\gamma_{\rm eff}(1)=0$,
i.e., adhesion increases the contact area even if no adhesion can be detected in a pull-off experiment.
Since it is the area of real contact which determines the sliding friction force, the adhesional
interaction may increase the sliding friction even if no adhesion can be detected in a pull-off experiment. 
We plan to study the adhesive contact between randomly rough surfaces using MD simulations to compare with the
predictions of the theory described above.

In an earlier publication one of us has studied
the effective binding energy as a function of the root-mean-square roughness for 7 differently
prepared surfaces\cite{Pere} (see Fig. \ref{Rubber.ball.Figure.3}). 
The theory curve (blue curve) in Fig. \ref{Rubber.ball.Figure.3} 
is based on the assumption
of perfect (atomic) contact at the interface. The
curves {\bf a} and {\bf b} are experimental results for the
pull-off velocity $0.2$ and $2 \ {\rm \mu m/s}$. The theory data point for 
$\gamma_{\rm eff} (1)/\Delta \gamma$ for
the highest roughness
value ($h_{\rm rms}=h_{\rm rms0} \approx 0.2 \ {\rm \mu m}$)
was actually negative (about $-0.1$, see Fig. \ref{adhesionfig}), indicating that the complete contact state 
cannot be the ground state, and in Fig. (\ref{Rubber.ball.Figure.3}) 
we therefore gave the binding energy $\gamma_{\rm eff}(1)=0$, of the fully detached state. 
However, in accordance with the experimental data
points, we find that with the improved elastic energy used above, the partly contact state
has a lower free energy (a larger binding energy) than the fully contact state, 
i.e., $\gamma_{\rm eff} (1)$ is positive
for the surface {\bf 7}, albeit still somewhat smaller than observed experimentally (compare Fig.
\ref{adhesionfig} for $h_{\rm rms}/h_{\rm rms0}=1$ with Fig. \ref{Rubber.ball.Figure.3} for 
$h_{\rm rms0}=0.2 \ {\rm \mu m}$).

\vskip 0.3cm

{\bf 10. Summary and Conclusion}

We have used our recently developed multiscale molecular dynamics approach
\cite{Chunyan} to study real contact area and interfacial separation from small contact to full contact.
The real contact area rises linearly with the load for small load. 
Here we have found that, at low magnification where the atomistic nature of the
solids becomes unimportant, the 
MD  results match very well with Persson's theory, 
especially when the contact approaches full contact. 

The interfacial separation as a function of squeezing pressure has been derived theoretically.
For non-adhesive interaction and small applied pressure, $p \propto {\rm exp}(-\bar{u}/u_{0})$ 
is in good agreement
with experimental observation. This relation has been tested with MD simulations and they match quite
well with each other. 
The present results may be of great importance for soft solids, 
e.g. rubber-like material, or very smooth surfaces.

\vskip 0.3cm

{\bf Acknowledgments}

We gratefully thank U. Tartaglino for many useful discussions related to the MD-simulation.
We thank G. Carbone for useful comments on the manuscript. 
We thank S. Hyun and M.O. Robbins for supplying the FEM data used in Fig. \ref{Rob1} and \ref{Rob2},
and the authors of Ref. \cite{Benz} for the height profile of the polymer surface 
in Fig. \ref{topo}.

\end{document}